\documentclass[prl,aps,floats,twocolumn,showpacs,nofootinbib]{revtex4-1}

\usepackage{amssymb}
\usepackage{amsmath}

\usepackage{graphicx}
\usepackage{graphics}
\usepackage{dcolumn}
\usepackage{color}
\usepackage{rotate}
\usepackage{fancyhdr} 
\usepackage{hyperref}
\usepackage{indentfirst}

\usepackage{umoline}
\usepackage{ulem}

\def\be{\begin{equation}}
\def\ee{\end{equation}}
\def\bea{\begin{eqnarray}}
\def\eea{\end{eqnarray}}
\def\ba{\begin{eqnarray}}
\def\ea{\end{eqnarray}}

\def\VEV#1{\left\langle #1 \right\rangle}

\newcommand{\sigmav}{\VEV{\sigma v}}
\newcommand{\vfo}{\VEV{v^{-1}}_{\rm fo}}

\definecolor{darkred}{rgb}{.743,0,0}

\begin{document}
\begin{flushright}UMD-PP-014-022 \\
\end{flushright}
\title{An Ultimate Target for Dark Matter Searches}

\author{Kfir Blum$^{1}$, Yanou Cui$^{2,3}$, and
     Marc Kamionkowski$^{4}$}
\affiliation{$^1$School of Natural Sciences, Institute for
Advanced Study, Princeton, New Jersey, 08540 USA\\
$^2$Perimeter Institute, 31 Caroline Street North
     Waterloo, Ontario N2L 2Y5 Canada \\
$^3$Maryland Center for Fundamental Physics, Department
     of Physics, University of Maryland, College Park, MD 20742, USA\\
$^4$Department of Physics and Astronomy, Johns
     Hopkins University, 3400 N.\ Charles St., Baltimore, MD
     21218, USA}
\date{\today}

\begin{abstract}
The combination of $S$-matrix unitarity and the dynamics of
thermal freeze-out for massive relic particles (denoted here simply by WIMPs) implies a {\it lower} limit on the density of such
particles, that provide a (potentially sub-dominant) contribution to dark
matter.  This then translates to lower limits to the signal rates
for a variety of techniques for direct and indirect detection of
dark matter.  For illustration, we focus on models where annihilation is
$s$-wave dominated.  We derive lower limits to the flux of
gamma-rays from WIMP annihilation at the Galactic center;
direct detection of WIMPs; energetic neutrinos from WIMP
annihilation in the Sun; and the effects of WIMPs on the angular
power spectrum and frequency spectrum of the cosmic microwave
background radiation.  The results suggest that a variety of dark-matter-search
techniques may provide interesting avenues to seek new physics,
even if WIMPs do not make up the total dark matter abundance.  While the limits
are quantitatively some distance from the reach of current
measurements, they may be interesting for long-range planning
exercises.
\end{abstract}
\pacs{95.35.+d,98.35.Gi}

\maketitle

Weakly-interacting massive particles (WIMPs) provide natural
dark matter (DM) candidates and may be experimentally
accessible.  This has led to much attention in the literature
(see, e.g.,
Refs.~\cite{Jungman:1995df,Bergstrom:2000pn,Bertone:2004pz}).
DM WIMPs are being sought directly in low-background detectors
\cite{Goodman:1984dc,Wasserman:1986hh}, indirectly
through searches for gamma rays, cosmic-ray positrons and
antiprotons produced by WIMP annihilation in the Galactic halo,
and through searches for energetic neutrinos from WIMP
annihilation in the Sun and Earth
\cite{Silk:1985ax,Kamionkowski:1991nj}.  Since these particles
arise from new physics beyond the Standard Model, evidence that dark
matter is composed of WIMPs would also comprise discovery of new
elementary particles.    Indeed, such particles are also sought
at the Large Hadron Collider (LHC).

There is always the possibility, though, that some new stable
massive particle exists but only constitutes a subdominant
component of DM \cite{Duda:2002hf}.  If the particle
interacts strongly enough to have been in thermal equilibrium
with the Standard Model plasma in the early Universe, then it
will still have some nonzero relic density today.  Such thermal
relic particles could thus show up in DM searches, even
if something else constitutes the majority of the dark
matter. Below we refer to stable massive particles with relic
abundance dictated by thermal freeze-out broadly as ``WIMPs'', 
keeping the name for simplicity even in cases where the particle
interacts strongly.

In this paper, we show that the combination of $S$-matrix
unitarity with the dynamics of thermal freeze-out in the early
Universe provides {\it lower} limits to the rates for detection
of WIMPs that make up a subdominant, and possibly negligible,
contribution to the total DM mass density.  Unitarity
provides an upper limit to annihilation cross sections, and this
has been used to derive an upper limit to the dark matter mass
\cite{Griest:1989wd} and upper limits to annihilation rates and
detection rates for dark matter in the current Universe
\cite{Beacom:2006tt,Hui:2001wy}. Still, it is somewhat
counterintuitive to think that unitarity can also provide {\it
lower} limits to detection rates. This conclusion, however,
follows simply because relic densities are inversely
proportional to the WIMP annihilation cross section and so,
given the upper limit to that cross section, bounded from below.
This then implies lower limits we derive to annihilation rates
in the Galactic halo (and thus---given particular final states
in the annihilation---to the fluxes of gamma rays, positrons, 
and antiprotons) and to rates for direct
detection and to fluxes of energetic neutrinos from WIMP
annihilation in the Sun/Earth. We also derive lower limits to
energy deposition from WIMP annihilation in the early Universe, leading to
changes to the cosmic microwave background (CMB) angular
power spectrum and to the amplitude of CMB spectral distortions. 
These limits may be valuable in the discussion of long-term
goals for the corresponding experimental avenues.

The relic density of a WIMP $\chi$ is $\Omega_\chi h^2 \simeq
0.1\,(\sigmav_0/\sigmav_{\rm fo})$, where $\Omega_\chi$ is the
fraction of the critical density contributed by the WIMP today,
$h\simeq0.7$ the Hubble parameter, $\sigmav_{\rm fo}$ the
thermally averaged velocity-weighted cross section for WIMP
annihilation (to all channels), calculated  at the time of
freeze-out, and the constant
$\sigmav_0=3\times10^{-26}$~cm$^3$~sec$^{-1}$ arises from the
dynamics of thermal freeze-out. 
For a pair of non-relativistic WIMPs annihilating with relative velocity $v$, partial-wave unitarity dictates an upper bound \cite{Griest:1989wd},
$\sigma_J \leq 4\pi (2J+1)/(m_\chi^2v^2)$,
where $m_\chi$ is the WIMP mass and $\sigma_J$ is the partial
cross-section for reaction with total angular momentum $J$. In
what follows we focus on the case where WIMP annihilation is
$s$-wave, or $J=0$. We then have 
\begin{equation}
     \sigmav_{\rm fo} \leq 4\pi\vfo/m_\chi^2,
\label{eqn:sigmalimit}
\end{equation}
where $\vfo=\sqrt{m_\chi/(\pi T_{\rm fo})}\simeq2.5$ is the
thermally averaged inverse relative velocity, using the
typical value $m_{\chi}/T_{\rm fo}=20$. 
There then follows a lower limit,
\begin{equation}
     \Omega_\chi/\Omega_{\rm dm}  \geq
     \left(m_\chi/110\, {\rm TeV} \right)^2,
\label{eqn:Omegalimit}
\end{equation}
to the relic density of WIMPs, in units of the observed DM density
$\Omega_{\rm dm}$, and where the numerical value is updated from
Ref.~\cite{Griest:1989wd} using the current value $\Omega_{\rm
dm}h^2\simeq 0.11$ \cite{Hinshaw:2012aka,Ade:2013zuv}.  The
usual unitarity limit $m_\chi\leq 110$~TeV to the WIMP mass
follows from the requirement $\Omega_\chi \leq \Omega_{\rm dm}$.

We now consider gamma rays from DM annihilation in the
halo of the Milky Way.  The search for such gamma rays is actively under
way; it is one of the principal science goals of the Fermi
Telescope \cite{Ackermann:2011wa, FermiLAT:2012aa} and will also be a target for
the Cherenkov Telescope Array (CTA) \cite{Vercellone:2014qia}.
The annihilation rate density is
\begin{equation}
     Q_\chi = \rho_\chi^2 \sigmav_h/(4 m_\chi^2),
\label{eqn:Qchi}
\end{equation}
where $\rho_\chi$ is the WIMP mass density and $\sigmav_h$ is the
velocity averaged annihilation cross section times relative velocity in the
Galactic halo. (If $\chi$ is self-conjugate, then the factor of $1/4$ on the RHS above should be replaced by $1/2$.) If $\chi$ constitutes 
a fraction $\Omega_\chi/\Omega_{\rm dm}$ of the DM,
then its density in the Galactic halo will be
$(\Omega_\chi/\Omega_{\rm dm})\rho_h$, where $\rho_h$ is the
Galactic-halo density.  For the $s$-wave annihilations we
consider here, neglecting effects such as Sommerfeld enhancement or suppression, $\sigmav_{\rm fo} = \sigmav_h$.  Then, using
$\Omega_\chi/\Omega_{\rm dm} \approx\sigmav_0/\sigmav_{\rm fo}$ and
Eq.~(\ref{eqn:sigmalimit}), we find a lower limit,
\begin{equation}\label{eq:Q}
    Q_\chi \geq \frac{\rho_h^2 \left(\sigmav_0\right)^2}{16\, \pi \vfo},
\end{equation}
independent of $m_\chi$ and $\langle\sigma v\rangle_{\rm fo}$ up to logarithmic corrections.

The differential gamma-ray flux from a window of solid angle
$\Delta\Omega$ around a given line of sight is
\begin{equation}
     J_\gamma(E_\gamma) = \int_{\Delta\Omega}
     \frac{d\Omega}{4\pi} \int\, dr\, Q_\chi(r)
     \frac{dN}{dE_\gamma},
\end{equation}
where the integral is along the line of sight, $Q_\chi(r)$ is
evaluated at a distance $r$ along that line of sight, and
$dN/dE_\gamma$ is the differential number of photons of energy
$E_\gamma$ per annihilation event.  Using Eq.~(\ref{eq:Q}), we have
\begin{eqnarray}
     J_\gamma(E_\gamma) &\geq& \frac{\bar
     J}{64 \pi^2\left(\sigmav_0\right)^2\vfo}
     \frac{dN}{dE_\gamma}  \nonumber \\
     & \simeq & 10^{-16} \left( 
     \frac{dN}{dE_\gamma} \right) \left(\frac{\bar J}{\bar
     J_{\rm nfw,gc}}\right)\, {\rm cm}^{-2}~{\rm sec}^{-1},
\end{eqnarray}
where $\bar J = \int_{\Delta\Omega} d\Omega \int\, dr\,
\rho_h^2$ is the line-of-sight integral. We have evaluated this
quantity in the second line in terms of the value $\bar J_{\rm
nfw,gc} \simeq 2.5\times 10^{21}$~GeV$^2$~cm$^{-5}$ obtained for
the HESS Galactic-center region of interest
\cite{Abramowski:2011hc} (a circle of radius $1^\circ$ around
the Galactic center with a Galactic-plane mask to remove
$|b|<0.3^\circ$) using the NFW profile
\cite{Navarro:1995iw}
$\rho_h(r)=\rho_0(r_s/r)\left(1+r/r_s\right)^{-2}$, with
$\rho_0=0.4$~GeV~cm$^{-3}$ and $r_s=20$~kpc. Besides the
Galactic center, another target of interest in gamma-ray
searches for DM are Milky Way dwarf galaxies, where $\bar J/\bar
J_{\rm nfw,gc}\sim10^{-4}-10^{-3}$ (see, e.g.,
Ref.~\cite{Abramowski:2010aa}) but astrophysical backgrounds are
smaller.

\begin{figure}[htbp]
\includegraphics[width=\linewidth]{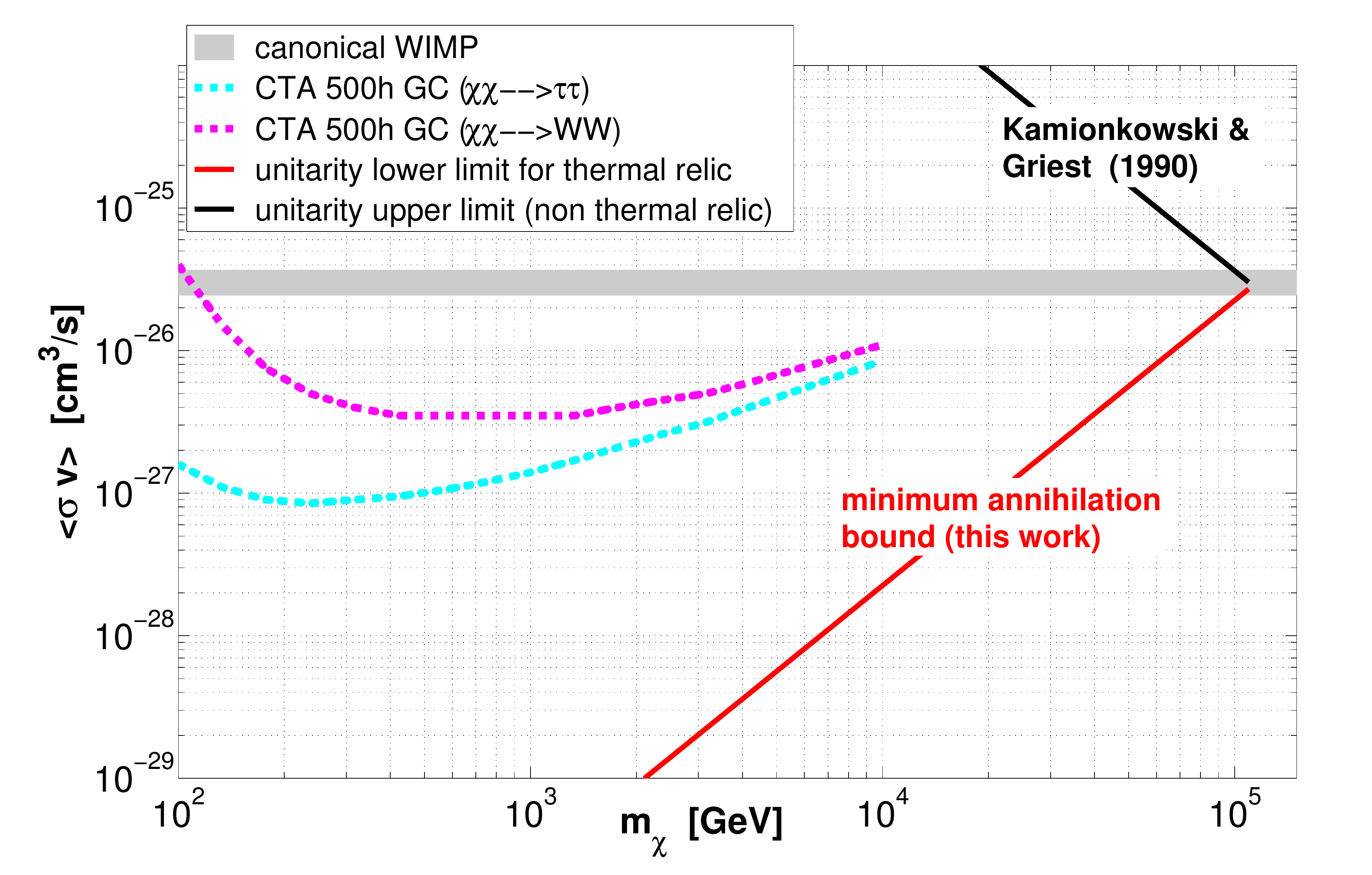}
\caption{Thermal averaged cross section times relative velocity $\sigmav$ versus
     WIMP mass $m_\chi$.  The horizontal gray band shows the
     canonical cross section for a thermal relic making up the dark matter.  The black line is the largest
     annihilation cross section consistent with unitarity.  The
     purple and cyan curves show an estimate of the smallest $\sigmav$
     detectable by CTA with 500 hours of observation
     time~\cite{Wood:2013taa}, assuming annihilation to
     $\tau^+\tau^-$ or $W^+W^-$ pairs, and assuming
     that the WIMP makes up all the halo dark matter.  The red
    line shows the smallest inferred $\sigmav$ that would be
     possible for a subdominant WIMP.}
\label{fig:gamma}
\end{figure}

We can assess the implications of the lower limit to the
gamma-ray flux of a thermal relic by projecting it
onto the sensitivity plot for gamma-ray experiments.  These
experiments typically show curves of $\sigmav$ versus $m_\chi$,
where the $\sigmav$ plotted is the value inferred from a given
gamma-ray flux, assuming that the WIMP comprises {\it all} the
DM.  In other words, $\sigmav$ is a proxy for (and
proportional to) the gamma-ray flux $J_\gamma$; i.e., $\sigmav
\propto \bar J_\gamma m_\chi^2$ [cf., Eq.~(\ref{eqn:Qchi})].

Fig.~\ref{fig:gamma} shows such a plot.  Shown as a gray
horizontal band is the value $\sigmav_0$ that arises if the WIMP
makes up all the DM.  The black line indicates the upper limit on $\sigmav$ directly imposed by unitarity (eq.(\ref{eqn:sigmalimit})).  The cyan and purple
curves show the smallest flux detectable with CTA 500 hours observation, under two
different assumptions about annihilation products.  
The red line indicates the smallest flux
possible for a subdominant thermal relic.  Clearly,
there is plenty of parameter space that gives rise to
gamma-ray fluxes well below those that will be accessible with
CTA. However, assuming that the annihilation final states include charged 
Standard Model fermions or massive gauge bosons, as in Fig.~\ref{fig:gamma}, 
subsequent generations of detectors that improve
the sensitivity sufficiently will be guaranteed to see a
gamma-ray signal if there is any stable thermal relic from the big
bang, even if that relic does not account for the majority of the DM.
For example, a telescope with a sensitivity improvement over CTA
of three orders of magnitude will see a signal from any thermal
relic heavier than about a TeV.

We now move on to direct detection of WIMPs.  The precise
expression for the rate for direct detection of WIMPs depends on
a variety of factors, including the DM velocity
distribution in the Galactic halo and energy dependence of the
WIMP-nucleus elastic-scattering cross section. If we
approximate the halo DM velocity distribution as a
Maxwell-Boltzmann distribution, the event rate per unit mass in a
DM detector is \cite{Jungman:1995df} $\Gamma =
2/\sqrt{\pi} \rho_\chi v_0 \sigma_N/(m_\chi m_N)$,
where $v_0\simeq 220$~km~sec$^{-1}$ is the halo circular speed,
$\sigma_N$ is the cross section for elastic scattering of the
WIMP from the nucleus, and $m_N$ the target-nucleus mass. The cross section for WIMP scattering off a nucleus of mass number
$A$ is related to the WIMP-nucleon cross section $\sigma_{\chi p (n)}$. For
instance, assuming spin-independent (SI) interaction without
isospin violation, the relation is $\sigma_N = (A^2\mu^2_{\chi
p}/\mu_{\chi N}^2) \sigma_{\chi p}$, where $\mu_{\chi p}$ and
$\mu_{\chi N}$  are the reduced mass of $\chi$-proton and
$\chi$-nucleus system respectively.  Replacing $\rho_\chi$ by the unitarity limit 
$\rho_h (\Omega_\chi/\Omega_{\rm dm}) \geq \rho_h
(m_\chi/110\,{\rm TeV})^2$, we infer that the rate for detection
of a WIMP with elastic cross section $\sigma_N$ must satisfy,
\begin{equation}
     \Gamma\geq \frac{2}{\sqrt{\pi}} \frac{\rho_h v_0 \sigma_N}
     {m_\chi m_N} \left(\frac{m_\chi}{110\, {\rm TeV}} \right)^2.  \label{eq:didt}
\end{equation}

\begin{figure}[htbp]
\includegraphics[width=\linewidth]{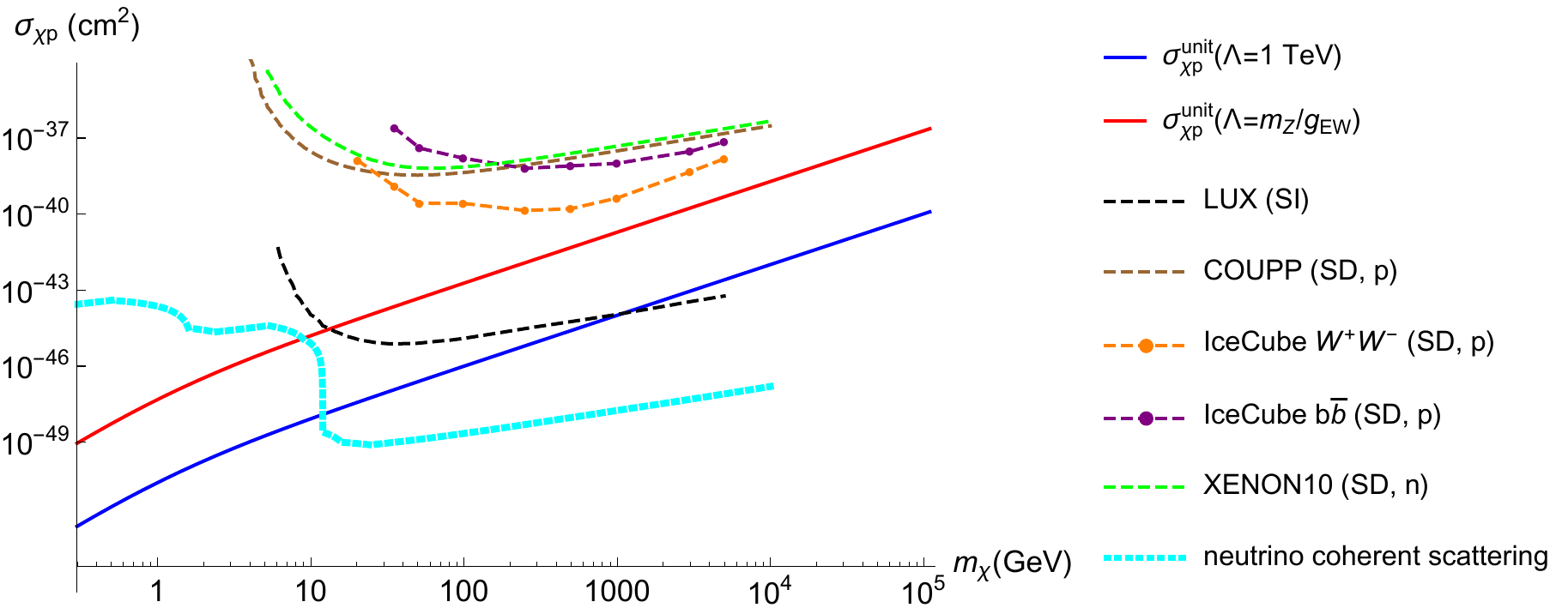}
\caption{WIMP-proton elastic-scattering cross section versus
     WIMP mass $m_\chi$. Dashed lines show the current lower limits by various experiments, 
     assuming that $\chi$ makes up all the
     dark matter. Solid blue and red lines denote the minimal effective
     WIMP-nucleon cross sections inferred from the unitarity limit, for two different values of the actual scattering cross section $\sigma_{\chi p}^{\rm actual}=\mu_{\chi p}^2/\pi\Lambda^4$, corresponding to $\Lambda=1$~TeV and $\Lambda\simeq m_Z/g_{\rm EW}$, respectively. The dotted cyan band shows the effective cross 
    section at which coherent scattering from background neutrinos becomes significant \cite{Cushman:2013zza}.}
\label{fig:direct}
\end{figure}

Again, the sensitivity of direct DM searches are
usually shown as plots of the WIMP-proton scattering
cross section $\sigma_{\chi p}$ versus WIMP mass $m_\chi$ (with an additional constraint for $\sigma_{\chi n}$ for spin-dependent (SD) interaction).  
These constraint plots then show the largest such cross section  allowed based on a given experiment,
assuming that the WIMP makes up all the DM. In Fig.~\ref{fig:direct} we show the smallest nominal
cross section $\sigma_{\chi
p}^{\rm unit}$ for a subdominant WIMP, obtained from our unitarity
argument, that would be inferred in this way, for different values of \textit{actual} scattering cross section $\sigma^{\rm actual}_{\chi
p}$.  According to Eq.(\ref{eq:didt}), we have the relation
$\sigma_{\chi p}^{\rm unit}\simeq (m_\chi/110\, {\rm TeV})^2
\sigma^{\rm actual}_{\chi p}$. The actual scattering cross
section $\sigma^{\rm actual}_{\chi
p}$, both SI and SD, in general can be parametrized by an
effective mass scale $\Lambda$ with $\sigma^{\rm actual}_{\chi
p}=\mu_{\chi p}^2/(\pi\Lambda^4)$. Note that at large WIMP mass
$m_\chi\gg m_p$,  $\sigma^{\rm actual}_{\chi
p}$ is independent of $m_\chi$ for a fixed $\Lambda$. In
particular, we show the example of a Standard Model $Z$ boson as
the mediator, i.e., $\Lambda\simeq m_Z/g_{\rm EW}$. In
Fig.~\ref{fig:direct}, together with the unitarity-limit lines,
we also show the current (lower) limits on SI scattering from
LUX, as well as limits on SD scattering from IceCube, XENON10 and COUPP
\cite{Angle:2008we,Behnke:2012ys,Aartsen:2012kia,Akerib:2013tjd}.

Limits from IceCube are directly based on energetic neutrinos
from WIMPs that are captured and then annihilate in the Sun, and
therefore depend on final states and on the equilibration between
the capture and annihilation processes, where the 
annihilation rate is equal to the half of the capture rate. This equilibration occurs, though, only if the equilibration
timescale \cite{Koushiappas:2009ee},
\begin{eqnarray}
     \tau &=& 1.6\times10^5\,{\rm yr}\,\left[\rho_{\chi,0.4} \sigmav_{26}f(m_\chi)\right]^{-1/2} \nonumber\\
     & & \times (m_\chi/100\,{\rm GeV})^{-3/4}
     \sigma_{40}^{-1/2},
\label{eqn:equilibration}
\end{eqnarray}
is shorter than the age of the Sun, $\sim 5\times 10^9$~yr. Here, $\rho_{\chi,0.4}$ is the WIMP density in units of
0.4~GeV~cm$^{-3}$, $f(m_\chi)\sim O(1)$ is given in
Ref.~\cite{Kamionkowski:1994dp}, $\sigmav_{26}$ the
annihilation cross section times relative velocity in units of
$10^{-26}$~cm$^3$~sec$^{-1}$, and $\sigma_{40} = \sigma_{\chi
p}/(10^{-40}\, {\rm cm}^2)$. As Eq.~(\ref{eqn:equilibration})
indicates, the equilibration timescale increases if the halo WIMP
density $\rho_\chi$ decreases, or if the annihilation cross
section $\sigmav_{26}$ decreases.  However, for
a thermal relic WIMP, the combination
$\rho_{\chi,0.4} \sigmav_{26}$ that appears in
Eq.~(\ref{eqn:equilibration}) remains constant as the
annihilation cross section (and thus relic density) is changed.
As a consequence, the energetic-neutrino flux for these
subdominant WIMPs is indeed controlled by the elastic-scattering
cross section, as long as $\sigma_{\chi p} \gtrsim (m_\chi/\rm 100~GeV)^{-3/2}10^{-49}~\rm cm^2$. 

We now turn to the effects of subdominant WIMP annihilation on
CMB fluctuations and spectral
distortions.  WIMP annihilation continuously injects a
small amount of energy into the cosmic plasma throughout the
history of the Universe.  Annihilations in the redshift range of
roughly $z\sim 10^3-10^6$ heat the plasma during a time when
photons cannot fully re-equilibrate thermally, giving rise to distortions in the CMB frequency
spectrum \cite{Chluba:2011hw}.  Annihilations that occur around
the time of
recombination alter slightly the ionization history of the
Universe and thus the detailed angular power spectrum of CMB
temperature and polarization fluctuations
\cite{Padmanabhan:2005es}.

The quantity of interest for CMB analysis is the rate density,
$\dot u = \rho_\chi^2 \sigmav/(2 m_\chi)$,
for energy injection due to WIMP annihilation, where here
$\rho_\chi$ is the cosmic density of WIMP $\chi$ at any particular
redshift. Current measurements of CMB anisotropy power spectra
imply an upper limit \cite{Hinshaw:2012aka,Ade:2013zuv},
\begin{equation}
     \left(\frac{\rho_\chi}{\rho_{\rm dm}} \right)^2 \left(
     \frac{ f \sigmav}{\sigmav_0} \right) \left
     (\frac{m_\chi}{10\, {\rm GeV}} \right)^{-1} \lesssim 1,\label{eq:CMB0}
\end{equation}
or equivalently,
\begin{equation}
     \dot u f \lesssim \dot u_{\rm ion,max} = \rho_{\rm
     dm}^2 \sigmav_0/(20\,{\rm GeV}).
\end{equation}
Here, the quantity $f\sim
0.1-1$ parametrizes the fraction of the injected energy that
gets absorbed by the plasma 
\cite{Chen:2003gz,Slatyer:2012yq}. The precise value of $f$
depends on the particular WIMP annihilation channels and also to
some extent on $m_\chi$. Assuming $\sigmav \simeq \langle\sigma v\rangle_{\rm fo}$, unitarity implies a lower limit,
\begin{equation}
     \dot u \geq m_\chi \rho_{\rm dm}^2 (\sigmav_0)^2/(8\pi \vfo),
\label{eqn:dotu}
\end{equation}
where $\rho_{\rm dm}(z) = \Omega_{\rm dm} \rho_c (1+z)^3$ is the
cosmic DM density at redshift $z$. Using Eq.~(\ref{eqn:dotu}), we have,
\begin{eqnarray}
     \frac{\dot u f}{\dot u_{\rm ion,max}} &\geq& \frac{5\, f
     \sigmav_0 m_\chi\,{\rm GeV}}{2\pi \vfo} \nonumber \\
     &\simeq& 2\times
     10^{-5}\, \left(\frac{f}{0.3} \right) \left(\frac{m_\chi}{100\,
     {\rm TeV}} \right).
\end{eqnarray}

Energy injection at redshifts $5\times 10^4 \lesssim z \lesssim
2\times 10^6$ (the $\mu$ era) give rise to $\mu$ distortions to the CMB
frequency spectrum, and injection at redshifts $1000\lesssim
z\lesssim 5\times10^4$ (the $y$ era), give rise primarily to Compton-$y$
distortions.  The current limits from the COBE/FIRAS~\cite{Fixsen:1996nj} experiment
are $|\mu| \lesssim 10^{-4}$ and $|y| \lesssim 10^{-5}$. Future
measurements by PIXIE~\cite{Kogut:2011xw} should reach a sensitivity of $\mu,y \sim
10^{-8}$ at $5\sigma$, and PRISM~\cite{Andre:2013afa} could get to values several
order of magnitude smaller. Energy injection during the $\mu$
era that changes the thermal energy density in the plasma by a
fractional amount $(\Delta \rho_\gamma/\rho_\gamma)$  
gives rise to a $\mu$ distortion of magnitude
 $\mu \simeq 1.4  (\Delta \rho_\gamma/\rho_\gamma)$, while that
 during the $y$ era gives rise to $y\simeq 0.25 (\Delta
 \rho_\gamma/\rho_\gamma)$ \cite{Chluba:2011hw}.  Annihilation during
 the $\mu$ era thus lead to \cite{Chluba:2011hw},
\begin{equation}
     \mu \simeq 4\times 10^{-8}\,(1-f_\nu)
     \left(\frac{\rho_{\chi}}{\rho_{\rm dm}}\right)^2 \left(
     \frac{\sigmav}{\sigmav_0} \right) \left(\frac{10\, {\rm
     GeV}}{m_\chi} \right),
\end{equation}
where $f_\nu$ is the fraction of the annihilation energy carried away by neutrinos, ranging between zero to tens of percent for typical annihilation final states. We can then write a
lower bound based on unitarity,
\begin{equation}
     \mu \gtrsim 3\times 10^{-12} \, (1-f_\nu)\,\left(
     m_\chi/100\,{\rm TeV} \right),
\end{equation}
to the $\mu$ distortion. We note that detailed spectrum measurements may help to disentangle a DM annihilation signal from other cosmic sources of spectral distortions~\cite{Chluba:2013pya}. For comparison, the not precisely
adiabatic cooling of primordial gas, as well as the dissipation
of small-scale acoustic waves, give rise to $\mu$ distortions at
the level of $\mu\sim 10^{-8}$~\cite{Chluba:2011hw}.   

To recapitulate, rates for direct and indirect detection of
subdominant WIMPs all depend either linearly or quadratically on
their relic density.  A lower limit to the density of a
thermal relic is set by the upper limit imposed by unitarity to
its annihilation cross section.  Thus, under common assumptions
for the annihilation final states or for the WIMP-nucleon elastic scattering cross section, the signals expected from
subdominant WIMPs cannot be arbitrarily small. Here we
have discussed direct WIMP detection at low-background
dark matter detectors, gamma rays from WIMP annihilation at the
Galactic center, energetic neutrinos from WIMP annihilation in
the Sun, and the effects of WIMPs on the angular power spectra
and frequency spectrum of the CMB. There are likewise 
lower limits to the flux of cosmic-ray positrons and antiprotons
and to the effects of WIMP annihilation on 21-cm fluctuations
from the dark ages \cite{Furlanetto:2006wp,Natarajan:2009bm}.

There are, of course, some caveats. First
of all, we have focussed on WIMPs with $s$-wave
annihilation. Quantitative results may differ for
$p$-wave annihilation or for WIMPs with Sommerfeld enhancements,
but in either case there will be limits that remain. The limits
do not necessarily apply for WIMPs that have non-thermal cosmic
densities, primordial quantum number asymmetry, or if there was
a significant amount of post-freezeout entropy production.
If the main annihilation channel of the WIMP is into stable dark-sector states such as dark radiation \cite{Garcia-Cely:2013wda,cmb_dm}, 
the limits for indirect searches and CMB signals we derived here
will bear a branching-fraction suppression. There are also caveats, though, that may strengthen the bounds. For example, the unitarity argument we have used is 
conservative and may be made more restrictive for large classes
of WIMP models \cite{Nussinov:2014qva}. To close, though, it is
of interest that simple considerations lead to a fairly general
lower limit to the rates for detection of thermal relics, even
if they do not make up most of the dark matter.

\vspace{6 pt}
M.K. thanks J. Chluba for useful discussions. K.B. was supported
by the DOE grant DE-SC000998, the BSF, and the Bahcall Fellow
Membership. Y.C. is supported by Perimeter
Institute for Theoretical Physics, which is supported by the
Government of Canada through Industry Canada and by the Province of
Ontario through the Ministry of Research and Innovation. Y.C. is in
part supported by NSF grant PHY-0968854 and by the Maryland Center
for Fundamental Physics. MK was supported
by the John Templeton Foundation, the Simons Foundation, NSF
grant PHY-1214000, and NASA ATP grant NNX15AB18G.

\end{document}